\documentstyle[12pt,graphics]{article}
\begin{document}
\title{Analysis of $RGU$ Photometry in Selected Area 51}
\author{S.~Bilir$^{1}$\footnote{corresponding author}, S.~Karaali$^{1}$, and R. Buser$^{2}$}       
\maketitle
{\center
$^{1}$\.Istanbul University Science Faculty, 
Department of Astronomy and Space Sciences\\ 
34119 \.Istanbul-TURKEY \\ sbilir@istanbul.edu.tr\\
		 \and
$^{2}$Astronomisches Institut der Universit\"{a}t Basel, 
Venusstrasse 7\\ 4102 Binningen-SWITZERLAND\\[3mm]
}
\date{}

\abstract{
A low-latitude anticenter field ($l=189^{\circ}$, $b=+21^{\circ}$) is 
investigated by using the full calibration tools of $RGU$ photometry. The 
observed $RGU$ data are reduced to the standard system and the separation 
of dwarfs and evolved stars is carried out by an empirical method. Stars 
are categorized into three metallicity classes, i.e. $-0.25<[M/H]\leq+0.50$, 
$-1.00<[M/H]\leq-0.25$, and $[M/H]\leq-1.00$ dex, and their absolute magnitudes 
are determined by the corresponding colour-magnitude diagrams. The unusually 
large scattering in the two-colour diagrams is reduced by excluding 153 
extra-galactic objects, identifying them compared with the charts of Basel 
Astronomical Institute and University of Minnesota, and by the criterion and 
algorithm of Gaidos et al. [1]. The local logarithmic space density for giants, 
$D^{*}(0)=6.75$, lies within the local densities of Gliese and Gliese \& 
Jahreiss. The local luminosity function for the absolute magnitude interval 
$3<M(G)\leq7$ agrees with Hipparcos' better than Gliese's, whereas there is a 
considerable excess for the interval $7<M(G)\leq8$ relative to both luminosity 
functions. This discrepancy may be due to many reasons, i.e. cumulative 
catalogue errors, binarity etc.}

{\bf Key Words: \rm Galaxy: structure -- Stars: luminosity function -- 
Photometry: RGU} 

\bigskip

\section{Introduction}

Buser's [2] photographic $RGU$ system is a systematic work based on synthetic 
photometry (Buser [3]). Galactic fields can be investigated through the method 
given by Buser \& Fenkart [4], thus main-sequence stars can be separated into 
three categories, i.e. Thin Disk, Thick Disk, and Halo, with available absolute 
magnitudes and metallicities. Additionally the standardized catalogues for 14 
fields, properly selected from Basel Halo Program are recently used for 
constructing a new Galactic model (Buser et al. [5, 6]). The lack 
of calibration of evolved stars (sub-giants and giants) in the work of Buser 
\& Fenkart [4] is compensated by Buser et al. [7], but without giving a method 
for their separation.

While standard star count analysis provide a description of the present 
structure of the Galaxy, additional data such as kinematics and chemical 
abundance are required for understanding the formation and evolution of our 
Milky Way Galaxy. Although Buser's $RGU$ photometry provides the metallicity 
distribution of field stars, it does not have an index suitable to the surface 
gravity, hence the separation of dwarfs (main-sequence stars) and evolved stars 
requires an indirect method (Karaali [8], Ak et al. [9], Karata\c{s} et al. 
[10], Karaali et al. [11], and Karata\c{s} et al. [12]). 

The excess of the luminosity function for absolutely faint magnitudes, i.e. 
$M(G)>6$ mag, had been used as a clue for such a separation in many works 
(Fenkart [13-16]) and apparently bright stars (roughly $G<15-16$ mag) with 
$M(G)>6$ mag on the main-sequence had been adopted as evolved stars with 
correspondingly brighter absolute magnitudes. A few iterations are sufficient to 
obtain a luminosity function agreeable with the luminosity functions of Gliese 
[17] and Hipparcos (Jahreiss \& Wielen [18, JW]). 

The comparison of Basel and Minnesota charts revealed that there is a 
considerable number of extra-galactic objects in the star fields, which cause 
an excess in the density and luminosity functions (Bilir et al. [19]). Hence, 
we applied the same procedure to eliminate such objects in our 
field. It turned out that 153 sources are extra-galactic objects, galaxies, 
occupying different regions in the two-colour diagrams. By adopting the 
\emph{distances from the stellar locus\/} criterion and the algorithm of Gaidos 
et al. [1] (see also Newberg \& Yanny [20]) however, with slight modification and 
purpose (see Section 3 for detail). Thus, recovery of the local 
luminosity function as given by Gliese [17] and/or Hipparcos (JW) could be 
possible. Section 2 is devoted to observations, reductions, and standardizations. 
Two-colour diagrams are given in Section 3, where the identification of 
extra-galactic objects, by comparison of the charts of Basel Astronomical 
Institute and Minnesota University, as well as the treatment of statistical 
scatter of stellar colours by means of criterion and algorithm of Gaidos et al. 
[1] is carried out. The separation of dwarfs and evolved stars and their absolute 
magnitude determination, and density and luminosity functions are given in 
Sections 4 and 5 respectively. Finally, summary and conclusion is presented in 
Section 6. We hope to derive the nature and distribution of stars in this field 
by applying the full calibration tools of $RGU$ photometry.  

\section {Observations, Reductions, and Standardization}

The coordinates of the field with size 0.45 square degrees are $\alpha =
07^{h}~28^{m}$, $\delta =+29^{o}~55{'}$; $l=189^{o}$, $b=+21^{o}$ (1950).
1737 stars were measured by one of us (S.K.) in 1995 at the Basel Astronomical 
Institute down to a apparent magnitude of $G=19$ on each five plates for each band, 
i.e. $R$, $G$, and $U$. 50 stars photoelectrically measured by Purgathofer [21] 
with 19.00, 18.65, and 17.65 as faintest $U-$, $B-$, and $V-$ magnitudes have 
been used as standards and their $UBV$ data were transformed to $RGU$-system by 
means of Buser's [22] formulae. The corresponding faintest $R-$, $G-$, and $U-$ 
magnitudes are 16.98, 18.25, and 20.09 respectively. The mean catalogue errors 
are given in Table 1. The $(U-B, B-V)$ two-colour diagram for standards reveals 
a colour-excess of $E(B-V)=0^{m}.03$ which corresponds to $E(G-R)=0^{m}.04$ in 
$RGU$ (Buser [2]). This value is close to those of Schlegel et al. 
[23], $E(B-V)=0^{m}.064$, and Burnstein \& Heiles [24], $E(B-V)=0^{m}.06$, 
who used different methods for their derivations, however. The first one 
is a model value, whereas the second one is derived from iso-obscuring contours 
with scale $0.03$ mag.   

$\Delta G$ versus $(G-R)_{obs}$ in Fig. 1a give no indication for a colour-equation 
for $G$, whereas there is a linear relation between $\Delta R$ versus $(G-R)_{obs}$, 
i.e. $\Delta R=-0.11(G-R)_{obs}+0.14$ (Fig. 1b), and a step function between 
$\Delta U$ versus $(U-G)_{obs}$ (Fig. 1c) as follows, where $\Delta m$ (m = $G$, 
$R$, and $U$) is the difference between the standard ($m_{s}$) and observed 
($m_{obs}$) apparent magnitudes, and $(G-R)_{obs}$ and $(U-G)_{obs}$ are the 
observed colour-indices: $\Delta U=$ +0.08, 0.00, -0.01, and +0.05 for 
$(U-G)_{obs}\leq1.00$, $1.00<(U-G)_{obs}\leq1.65$, $1.65<(U-G)_{obs}\leq2.20$, 
and $2.20<(U-G)_{obs}$, respectively.

All the $RGU$ data are reduced to the standard system by applying the corrections 
mentioned above. Thus, all magnitudes and colours which will be used henceforth 
are dereddened and standard ones.   

\section{Two-Colour Diagrams} 
The two-colour diagrams are drawn within the limiting apparent magnitude of 
$G=18$ for consecutive $G$- apparent magnitude intervals, where four of them, 
i.e. (14.0-15.0], (15.5-16.0], (16.5-17.0], and (17.5-18.0] are given in 
Fig. 2, respectively, as examples. As cited in Section 1, there is an unusually 
large scatter in these diagrams for low latitude field ($b=+21^{o}$), especially 
in the location of metal-poor stars in apparently faint magnitude intervals. 
The comparison of the charts of the Basel Astronomical Institute and Minnesota 
University reveals that 153 of 1737 objects are extra-galactic ones. 
However, omitting these objects does not reduce the scattering considerably, 
because they lie even within the region occupied by stars, i.e. 
$-3.0<[M/H]<+0.5$ dex. The extra-galactic objects in the two-colour diagrams 
given in Fig. 2 are marked with a different symbol ($\triangle$).

The luminosity function for all stars (without extra-galactic objects) 
within the limiting apparent magnitude, $G=18$, (Fig. 5b) resulting from 
the comparison of density functions with model gradients (see Section 5) 
Buser et al. [5, 6] (henceforth, BRK) deviates systematically 
from that of Gliese [17], i.e. there is an excess of absolutely faint 
stars, $M(G)>6$, and deficience of absolutely bright stars, $M(G)<5$, 
indicating that the scattering affected the absolutely bright stars to 
shift to the region of faint ones. It is worth noting that this is what 
we had experienced in our other works (cf. Fenkart \& Karaali [25]).

We applied the \emph{distances from the stellar locus\/} criterion and the 
algorithm of Gaidos et al. [1] (see also Newberg \& Yanny [20]) with a slight 
modification and purpose, however, to reduce the number of scattered stars. 
These authors formed a locus of all pointlike sources in the multi-colour 
space and they fitted a set of locus points along the center of the locus 
of these sources. The stellar candidates selected were those that were closer 
to their associated locus point than the metric distance $d$ (a parameter to 
be determined) in magnitudes for all colours, whereas the quasi-stellar object 
candidates were the ones at distances larger than $d$. In our case, we applied 
this criterion and algorithm to the colour-plane, i.e. $(U-G, G-R)$ two-colour 
diagram, and adopted the metric distance as $d=1.3s$, where $s$ is the 
standard deviation for each colour, for each sub-sample of stars (separated by 
dashed lines in Fig. 3b). Thus, stars for each sub-sample, within at least 
$1s$ were included in the statistics (see Table 2 for their percentages). 
Fig. 3a gives all dwarfs in the field SA 51, and Fig. 3b those selected by the 
criterion and algorithm of Gaidos et al. [1]) for statistical purpose.

\section{Dwarf-Giant Separation and Absolute Magnitude Determination}

Dwarfs and late-type giants were separated by the gap-criterion (Becker [26]) 
for a long time, whereas no effort was carried out for the separation of 
sub-giants. Late-type giants were recognized by their location separated from 
dwarfs by a gap and with larger $U-G$ colour-indices relative to the 
main-sequence with $[M/H]\sim0.0$ dex, in the $(U-G, G-R)$ two-colour diagrams 
for low-latitude fields (cf. Becker \& Fang [27]; Hersperger [28]). However, 
Becker [29] showed that there exists another type of late-type giant, lying at 
the metal-poor region of the two-colour diagram, and a bit bluer than the ones 
mentioned above, thus a bit disregarding the gap which separates dwarfs and 
metal-rich late-type giants. During the epoch of comparison the density 
functions with the galactic models, the local logarithmic space density for 
late-type giants, i.e. $\odot=6.64$ (Gliese [17]), was the favour clue for 
their separation (Del Rio \& Fenkart [30]; Fenkart [13-16]; Fenkart \& Karaali 
[25]). 

Systematic deviation of the luminosity functions from the one of Gliese [17] 
revealed that the absolutely faint segment of the luminosity function was 
contaminated by evolved stars (sub-giants and giants), resulting in an excess 
for $M(G)>6$ mag and a deficience for $M(G)<5$ mag in the luminosity function. 
This disagreement was used as a clue for the separation of dwarfs and evolved 
stars in recent years (Karaali [8], Ak et al. [9], Karata\c{s} et al. [10], 
Karaali et al. [11], and Karata\c{s} et al. [12]). The fundamental assumption 
for this empirical method is that apparently bright and absolutely faint stars 
on the main-sequence are evolved. In this work, a few iterations provided a 
luminosity function in best agreement with the local luminosity function as 
given by Gliese [17] and/or Hipparcos (JW) by assuming that for apparent 
magnitudes brighter than $G=15.5$ mag, stars which according to their positions 
in the two-colour diagram could be identified as dwarfs with assigned absolute 
magnitudes fainter than $M(G)=6$ mag, are however most likely evolved stars 
with correspondingly brighter absolute magnitudes. 

Following Buser \& Fenkart [4]; we separated dwarfs into three metallicity-classes, 
i.e. $-0.25<[M/H]\leq+0.50$ dex (Thin Disk), $-1.00<[M/H]\leq-0.25$ dex (Thick 
Disk), and $[M/H]\leq-1.00$ dex (Halo), and we used their corresponding 
colour-magnitude diagram, derived from extent sources via synthetic photometry, 
for absolute magnitude determination. Contrary to the works investigated in 
Steinlin's [31] system, individual absolute magnitudes are adopted for late-type 
giants (and sub-giants) by separating them into different metallicity-classes and 
using the multi-metallicity colour-magnitude diagram of Buser et al. [7], derived 
in the same way as dwarfs.

\section{Density and Luminosity Functions}
The logarithmic space densities $D^{*}=logD(r)+10$ are evaluated for five 
absolute magnitude intervals, i.e. (3-4], (4-5], (5-6], (6-7], and (7-8], 
where the absolute magnitudes are complete, and for late-type giants 
(Tables 3 and 4). However, the number of stars for the absolute magnitude 
intervals (2-3], (8-9], and (9-10] for each distance interval is also given in 
Table 3. Here $D=N/\Delta V_{1,2}$, N being the number of stars, found in the 
partial volume $\Delta V_{1,2}$ which is determined by its limiting distances 
$r_{1}$ and $r_{2}$, and by the apparent field size in square degrees $A$, i.e. 
$\Delta V_{1,2}=(\pi/180)^{2}(A/3)(r_{2}^{3}-r_{1}^{3})$.

The density functions are most appropriately given in the form of histograms 
whose sections with ordinates $D^{*}(r_{1}, r_{2})$ cover the distance-intervals 
($r_{1}, r_{2}$), and heavy dots on the histogram sections $D^{*}(r_{1}, r_{2})$ 
designate the centroid-distance $\bar{r}=[(r^{3}_{1}+r^{3}_{2})/2]^{1/3}$ of the 
corresponding partial volume $\Delta V_{1,2}$ (Del Rio \& Fenkart [30]; Fenkart 
\& Karaali [25]; and Fenkart [13-16]).

The density functions are compared with the galactic model of BRK, in the form 
$\Delta log D(r)=logD(r, l, b)-logD(0, l, b)$ versus $r$, where $\Delta logD(r)$ 
is the difference between the logarithmic densities at distance $r$ and at the Sun. 
Thus, $\Delta log D(r)=0$ points out the logarithmic space density for $r=0$ which 
is available for local luminosity function determination. The comparison is carried 
out as explained in some works of Basel fields (Del Rio \& Fenkart [30]; Fenkart \& 
Karaali [25], i.e. by shifting the model curve perpendicular to the distance axis 
until the best fit to the histogram results at the centroid distances (Fig. 4).

Fig. 4 show that there is a good agreement between the model gradients 
and the observed density histograms. The same agreement holds when local densities 
are considered, except for the absolute magnitude interval $7<M(G)\leq8$. This can be 
confirmed by comparison of the local luminosity function with the luminosity function 
of Gliese [17] and Hipparcos (JW). In Fig. 5, there are two luminosity functions 
resulting from comparisons of observed density histograms for dwarfs and sub-giants 
with the best-fitting model gradients BRK. For (a) we used the data in Table 3 and 
Fig. 4a-e where unusual scattering in the two-colour diagrams is reduced by the criterion 
and algorithm of Gaidos et al. [1], whereas for (b) all dwarfs and sub-giants within 
the limiting apparent magnitude, $G=18$, are used. The agreement is much better for 
(a). The luminosity (a) for the interval $5<M(G)\leq6$, is almost equal to those of 
Gliese and Hipparcos and close to them for the interval $6<M(G)\leq7$, but it is a bit 
deficient relative to the luminosity function of Hipparcos for the segment 
$3<M(G)\leq5$ (the luminosity function of Hipparcos is also deficient relative to the 
luminosity function of Gliese for the same absolute magnitude interval). However, 
there is a considerable excess for the luminosity function (a) relative to both 
luminosity functions of Gliese and Hipparcos for the interval $7<M(G)\leq8$, i.e. 
$0.30$ in units of logarithmic space density which is much larger than the standard 
deviation for this absolute magnitude interval (Table 5). It is worth noting to note 
that the differences between the luminosity function (a) and that of Hipparcos for 
other absolute magnitude intervals are all less than the corresponding standard 
deviations given in Table 5. Although the luminosity function (b) is close to the 
luminosity function (a) for the absolute magnitude intervals (3-4], (4-5], and (5-6], 
it deviates from (a) for two absolutely faint magnitude intervals, i.e. (6-7], and 
(7-8], considerably.

The comparison of the density function for giants with the model gradients BRK 
is carried out up to $r=10$ kpc (Fig. 4f). Six stars within the large distance 
interval $10.00<r\leq19.95$ kpc are not included in the statistics. The local 
density resulting from this comparison, $D^{*}(0)=6.75$, lies between the local 
densities of Gliese [17] and Gliese \& Jahreiss [32], i.e. $\odot=6.64$ and 
$\odot=6.92$, respectively.                                                                           

\section{Summary and Conclusion}

We used the full calibration tools of $RGU$ photometry to investigate the 
low-latitude ($b=+21^{o}$) and anticenter ($l=189^{o}$) field SA 51. The 
observed $RGU$ data are reduced to the standard system and the separation 
of dwarfs and evolved stars is carried out by an empirical method based on 
the assumption that apparently bright stars are evolved (Karaali [8], Ak et 
al. [9], Karata\c{s} et al. [10], Karaali et al. [11], and Karata\c{s} 
et al. [12]), i.e. for apparent magnitudes brighter than $G=15.5$ mag, 
stars which, according to their positions, are identified as 
dwarfs with assigned absolute magnitude fainter than $M(G)=6$ mag, are 
however most likely evolved stars with corresponding brighter absolute 
magnitudes. This assumption provided a luminosity function agreeable with 
the local luminosity function as given by Gliese [17] and Hipparcos (JW). 
Dwarfs are separated into three metallicity classes, i.e. 
$-0.25<[M/H]\leq+0.50$ dex (Thin Disk), $-1.00<[M/H]\leq-0.25$ dex 
(Thick Disk), and $[M/H]\leq-1.00$ dex (Halo), and their absolute magnitudes 
are determined by the corresponding colour-magnitude diagrams of Buser \& 
Fenkart [4], derived from extent sources via synthetic photometry. The 
metallicities and absolute magnitudes for evolved stars are evaluated by 
the recent diagrams of Buser et al. [7]. 

Although 153 extra-galactic objects were excluded from the complete sample, 
compared with the charts of Basel Astronomical Institute and Minnesota 
University (Bilir et al. [19]), the scattering in the two-colour diagrams 
could not be reduced. We applied the criterion and algorithm of Gaidos 
et al. [1] to the colour-plane, i.e. $(U-G, G-R)$ two-colour diagram, to 
reject dwarfs at distances in magnitude larger than $d=1.3s$ from the 
center of the locus of all dwarfs in the direction to $U-G$ and $G-R$ 
axes, where $s$ is the standard deviation of dwarfs associated with the 
locus point in each sub-sample (separated by dashed lines in Fig. 3b). 
This limitation reduced dwarfs by $79\%$ which is larger than the 
percentage in $1s$ for a gaussian distribution.

The density histograms for dwarfs and sub-giants with absolute magnitudes 
(3-4], (4-5], (5-6], (6-7], and (7-8] agree with the model gradients BRK. 
The same agreement holds when local densities are considered, except for 
the absolute magnitude interval $7<M(G)\leq 8$ where the luminosity has an 
excess of $0.30$ in units of logarithmic density relative to the luminosity 
of Hipparcos. The number of dwarfs in this interval can not be reduced, 
otherwise they turn out to be giants with density function contradicting 
with the model gradients BRK and local density different than the ones of 
Gliese [17] and Gliese \& Jahreiss [32].  

One of the reasons for the deviation of the luminosity function for the 
interval $7<M(G)\leq8$ from Gliese's [17] or Hipparcos' (JW) may be binarity, 
besides others such as cumulative catalogue errors etc. We refer to Buser \& 
Kaeser [33], who were the first to consider the effects of unresolved stars 
in the far-field surveys and luminosity functions. It may require the 
comparison of the luminosity functions with an appropriately redetermined 
local one via the data of Gliese [17] or Hipparcos (JW). 

The luminosity function in our work is much better than the one in 
Karata\c{s} et al. [10]. All the tools used for the investigation 
of two fields are the same, except the distance criterion and algorithm of Gaidos 
et al. [1] which is used only in our work. This new approach can be useful for 
understanding the nature of stars in the fields treated.

Acknowledgements: We would like to thank Schweizerischer Nationalfonds zur 
F\"{o}rderung der Wissenschaflichen Forschung, and the Scientific and Technical 
Research Council of Turkey for financial support (TBAG-AY/74), and the University 
of Minnesota for providing us with its APS on-line catalogues for the 
investigated field.

\begin{figure}
%\vskip 12 cm
%\hskip4cm
%\special{bmp: c:fig01.bmp x=7.75cm y=11.75cm}
\resizebox{7.75cm}{11.75cm}{\includegraphics*{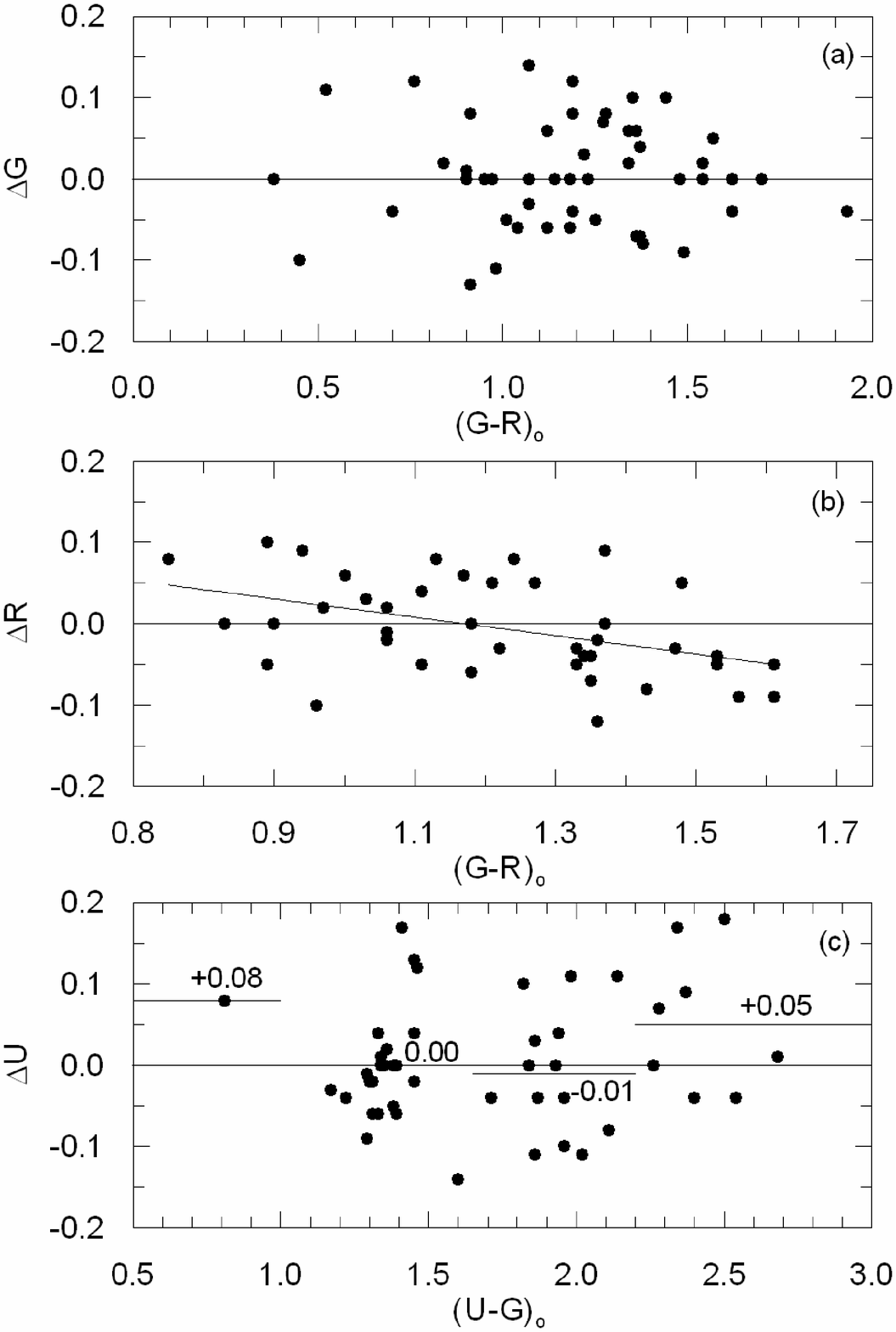}} 
\caption[] {Standardization of the data. There is no indication for any 
colour - equation for G (a), whereas there is a linear relation between 
$\Delta R$ versus $(G-R)_{obs}$ (b), and a step function between 
$\Delta U$ versus $(U-G)_{obs}$ (c).} 
\end {figure}

\begin{figure}
%\vskip 17 cm
%\hskip 1.75cm
%\special{bmp: c:fig02.bmp x=12.20cm y=16cm}
\resizebox{12.20cm}{16cm}{\includegraphics*{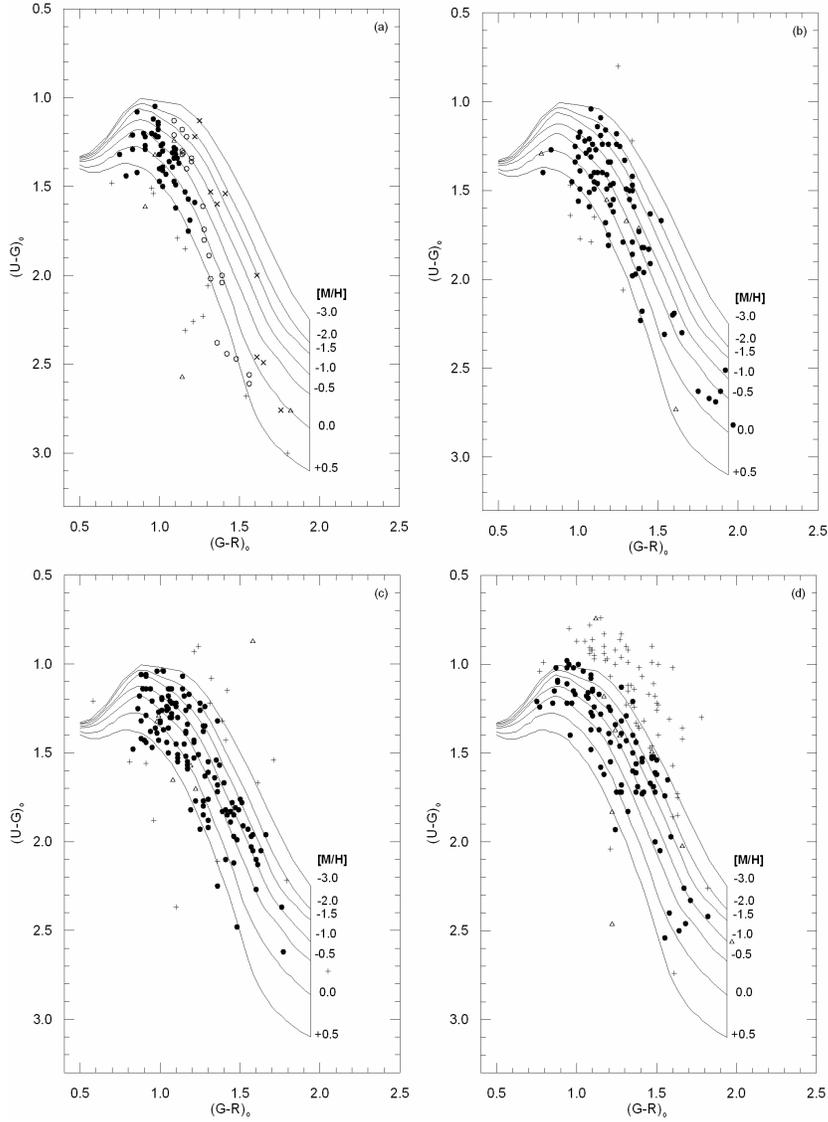}} 
\caption {Four two-colour diagrams as examples, for (14.0-15.0] (a), 
(15.5-16.0] (b), (16.5-17.0] (c), and (17.5-18.0] (d). 
There is an unusually large scatter, especially in the faintest apparent 
magnitude interval, in the location of metal-poor stars. Symbols: 
($\bullet$) dwarfs, (o) sub-giants, (x) late-type giants, ($\triangle$) 
extra-galactic objects, and (+) not included into statistics.}   
\end {figure}

\begin{figure}
%\vskip 9.50cm
%\hskip .5cm
%\special{bmp: c:fig03.bmp x=13cm y=9.45cm}
\resizebox{13cm}{9.45cm}{\includegraphics*{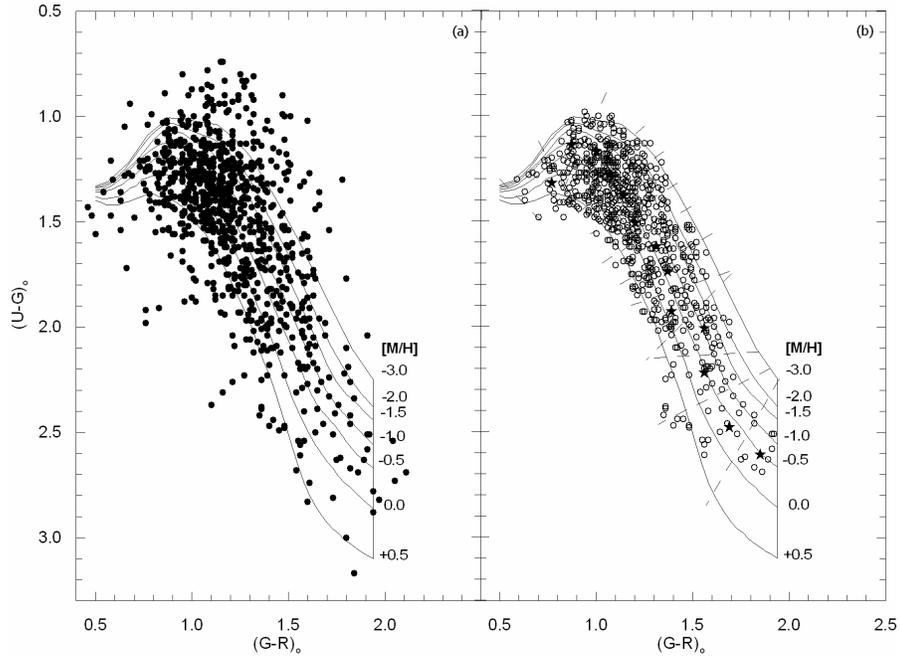}} 
\caption[] {Two-colour diagram for dwarfs brighter than the apparently 
limiting magnitude, $G=18$ mag. (a) for all stars, (b) for stars included 
into statistics, i.e. within the distance $d=1.3s$, for $U-G$, and $G-R$ 
colours, from the corresponding locus point associated. The symbol $(\star)$ 
denotes the locus point and the dashed lines separate dwarfs into sub-samples 
with centroid $(\star)$. $s$: standard deviation for each colour, for the 
corresponding sub-sample.}
\end {figure}

\begin{figure}
%\vskip 12 cm
%\hskip 1.5cm
%\special{bmp: c:fig04.bmp x=12.65cm y=12cm}
\resizebox{12.65cm}{12cm}{\includegraphics*{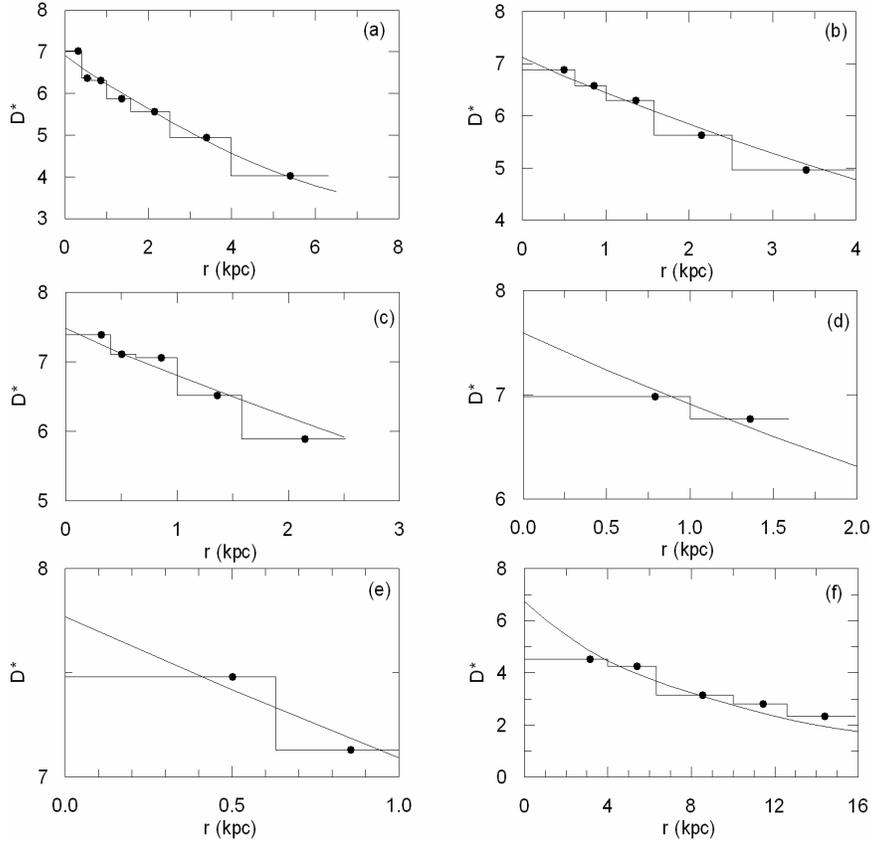}} 
\caption[] {Logarithmic space-density histograms for dwarfs and sub-giants of 
different absolute-magnitude intervals: (3-4] (a), (4-5] (b), (5-6] (c), 
(6-7] (d), (7-8] (e), and for late-type giants (f). ($\bullet$) centroid 
distance within the limiting distance of completeness, for comparison with 
model gradients BRK.}
\end {figure}

\begin{figure}
%\vskip 5cm
%\hskip 4.5cm
%\special{bmp: c:fig05.bmp x=8cm y=4.75cm}
\resizebox{8cm}{4.75cm}{\includegraphics*{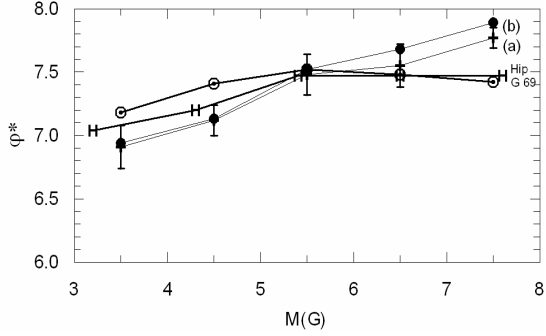}} 
\caption[] {Two luminosity functions resulting from comparison of observed 
histograms with best-fitting model gradients BRK, and confronted to Gliese's 
[17] ($\odot$), and Hipparcos' [18] (H). (a) for dwarfs and sub-giants for 
which unusually scattering in the two-colour diagrams is reduced by the 
criterion and algorithm of Gaidos et al. [1] (taken from Table 3), and (b) 
for all dwarfs and sub-giants in the two-colour diagrams within the limiting 
apparent magnitude, $G=18$ (the density functions for (b) have not been given 
to avoid space consuming).}
\end {figure}

\begin{table}
\caption[]{Mean catalogue errors.}
\begin{center}
\begin{tabular}{cccc}
\hline
$G$ interval  & $(G)_{err}$ & $(G-R)_{err}$ & $(U-G)_{err}$ \\
\hline
  $<$12 & $\pm$ 0.02 & $\pm$ 0.02 & $\pm$ 0.03 \\
(12-14] &       ~~~~0.02 &       ~~~~0.02 &       ~~~~0.03 \\
(14-16] &       ~~~~0.02 &       ~~~~0.03 &       ~~~~0.02 \\
  $>$16 &       ~~~~0.03 &       ~~~~0.04 &       ~~~~0.03 \\
\hline
\end{tabular}  
\end{center}
\end{table}

\begin{table}
\caption[]{The $U-G$, $G-R$ colour-indices of the locus points (W), number of 
stars, for each sub-sample, associated with them ($N^{'}$) and within the 
distance $d=1.3s$ from the corresponding locus point (N), and the percentage 
of stars included into statistics ($s$: standard deviation for each colour for 
the sub-sample considered).}
\begin{center}
\begin{tabular}{cccccc}
\hline
    W &      $U-G$ &       $G-R$&   $N^{'}$ &          N &       $\%$ \\
\hline
    1 &       1.32 &       0.77 &         30 &         20 &         67 \\
    2 &       1.14 &       0.87 &         50 &         41 &         82 \\
    3 &       1.17 &       1.00 &        107 &         76 &         71 \\
    4 &       1.28 &       1.07 &        146 &        133 &         91 \\
    5 &       1.38 &       1.14 &        144 &        108 &         75 \\
    6 &       1.51 &       1.20 &        100 &         81 &         81 \\
    7 &       1.62 &       1.31 &         67 &         53 &         79 \\
    8 &       1.74 &       1.37 &         81 &         62 &         77 \\
    9 &       1.93 &       1.39 &         53 &         41 &         77 \\
   10 &       2.01 &       1.56 &         38 &         29 &         76 \\
   11 &       2.22 &       1.56 &         25 &         18 &         72 \\
   12 &       2.48 &       1.69 &         22 &         19 &         86 \\
   13 &       2.61 &       1.85 &         10 &          8 &         80 \\
\hline
\end{tabular}
\end{center}
\end{table} 

\begin{table*}
\caption[]{The logarithmic space densities $D^{*}=\log D+10$ for five absolute 
magnitude intervals, i.e. (3-4], (4-5], (5-6], (6-7], and (7-8] for dwarfs and 
sub-giants, where the absolute magnitudes are complete. Thick horizontal lines: 
limiting distance of completeness (for definition of the symbols see text, 
distances in kpc, volumes in pc$^{3}$).}
\begin{center}
\scriptsize{
\begin{tabular}{cccrrrrrrrr}
\hline
 &    &    $M(G)\rightarrow$& (2-3]  & (3-4]    & (4-5]    & (5-6] & (6-7] & (7-8] & (8-9]  & (9-10] \\
\hline
$r_{1}$-$r_{2}$   & $\Delta V_{1,2}$ & $\bar{r}$&  N~~D*   &  N~~D* &  N~~D* & N~~D* & N~~D*  & N~~D* & N~~D*& N~~D*\\
\hline
0.00-0.40 & 2.88 (3) & 0.32 &      &  3~~7.02 &         &   7~~7.39 &         &         &    19~~-- & 4~~-- \\
0.00-0.63 & 1.15 (4) & 0.50 &      &          &  9~~6.89&           &         & 35~~7.48&           &       \\
0.00-1.00 & 4.57 (4) & 0.79 &      &          &         &           & 44~~6.98&         &           &       \\
0.40-0.63 & 8.60 (3) & 0.54 &      &  2~~6.37 &         &  11~~7.11 &         &         &    28~~-- & 3~~-- \\
0.63-1.00 & 3.42 (4) & 0.86 &      &  7~~6.31 & 13~~6.58&  39~~7.06 &         & 46~~7.13& 15~~-- & \\ \cline {9-9}
1.00-1.59 & 1.36 (5) & 1.36 &      & 10~~5.87 & 27~~6.30&  45~~6.52 & 80~~6.77& 24~~6.25 & \\ \cline {8-8}
1.59-2.51 & 5.42 (5) & 2.15 & 4~~-- & 20~~5.57& 23~~5.63&  42~~5.89 & 28~~5.71 & & & \\ \cline {7-7}
2.51-3.98 & 2.16 (6) & 3.40 & 4~~-- & 19~~4.94& 20~~4.97&  14~~4.81 & & & &  \\ \cline {6-6}
3.98-6.31 & 8.59 (6) & 5.40 & 1~~-- &  9~~4.02&  7~~3.91&           & & & &  \\ \cline {5-5}
\\
\hline
          &        &Total & 9~~~~   & 70~~~~~~~~  & 99~~~~~~~& 158~~~~~~~ & 152~~~~~~~ & 105~~~~~~~ & 62~~~~& 7~~~~ \\
\hline
\end{tabular}  
}
\end{center}
\end{table*}

\begin{table}
\caption[]{The logarithmic space density $D^{*}= \log D + 10$ for late-type giants 
(see text for definition of the symbols, distances in kpc, volumes in pc$^{3}$)}
\begin{center}
\begin{tabular}{ccccc}
\hline
$r_{1}$-$r_{2}$ & $\Delta V_{1,2}$ &$\bar{r}$ & N &$D^{*}$\\
\hline
     0-3.98& 2.88 (6) &  3.16 & 10 & 4.54 \\
  3.98-6.31& 8.60 (6) &  5.40 & 16 & 4.27 \\
 6.31-10.00& 3.42 (7) &  8.55 &  5 & 3.16 \\
10.00-19.95& 3.17 (8) & 16.48 &  6 & $--$ \\
\hline
\end{tabular}  
\end{center}
\end{table}

\begin{table}
\caption[]{Local luminosity function resulting from comparison of observed 
histograms with best-fitting model gradients BRK, and confronted to Gliese's 
[17] and Hipparcos' (JW) local luminosity functions. $s$: standard deviation 
in units of logarithmic space density.}
\begin{center}
\begin{tabular}{cc}
\hline
$M(G)\rightarrow$&(3-4]~~~(4-5]~~~(5-6]~~~(6-7]~~~(7-8]\\
\hline
BRK           & 6.91~~~7.12~~~~7.48~~~~7.55~~~~7.77\\
    $s(\pm)$  & 0.17~~~0.12~~~~0.16~~~~0.17~~~~0.08\\
Gliese (1969) & 7.18~~~7.41~~~~7.52~~~~7.48~~~~7.42\\
Hipparcos (JW)& 7.04~~~7.20~~~~7.47~~~~7.47~~~~7.47\\
\hline
\end{tabular}  
\end{center}
\end{table}


\begin{thebibliography}{}
\bibitem{} E. J. Gaidos, E. A. Magnier, \& P. L. Schechter, \it PASP, \bf 105, 
  \rm (1993), 1294. 
\bibitem{} R. Buser, \it A\&A, \bf 62, \rm (1978), 425.
\bibitem{} R. Buser, PhD. Thesis, Astronomisches Institut der Universit\"{a}t Basel, 
  Switzerland, 1975.
\bibitem{} R. Buser, \& R. P. Fenkart, \it A\&A, \bf 239, \rm (1990), 243.
\bibitem{} R. Buser, J. Rong, \& S. Karaali, \it A\&A, \bf 331, \rm (1998), 934 (BRK).
\bibitem{} R. Buser, J. Rong, \& S. Karaali, \it A\&A, \bf 348, \rm (1999), 98.
\bibitem{} R. Buser, Y. Karata\c{s}, Th. Lejeune, J. X. Rong, P. Westera, 
  \& S. G. Ak, \it A\&A, \bf 357, \rm (2000), 988.
\bibitem{} S. Karaali, VIII. Nat. Astron. Symp. Eds. Z. Aslan  \& O. 
  G\"{o}lba\c{s}\i, Ankara-Turkey, (1992), p.202.
\bibitem{} S. G. Ak, S. Karaali, \& R. Buser, \it A\&AS, \bf 131, \rm (1998), 345.
\bibitem{} Y. Karata\c{s}, S. Karaali, \& R. Buser, \it A\&A, \bf 373, 
  \rm (2001), 895
\bibitem{} S. Karaali, S. Bilir, \& R. Buser, \it PASA\rm, (2004) (accepted).
\bibitem{} Y. Karata\c{s}, S. Bilir, S. Karaali, \& S. G. Ak, \it AN\rm,  
  (2004) (accepted).
\bibitem{} R. P. Fenkart, \it A\&AS, \bf 78, \rm (1989), 217.
\bibitem{} R. P. Fenkart, \it A\&AS, \bf 79, \rm (1989), 51.
\bibitem{} R. P. Fenkart, \it A\&AS, \bf 80, \rm (1989), 89.
\bibitem{} R. P. Fenkart, \it A\&AS, \bf 81, \rm (1989), 187.
\bibitem{} W. Gliese, Ver\"{o}ff. Astron. Rechen Inst. Heidelberg, (1969), No:22.
\bibitem{} H. Jahreiss, \& R. Wielen, in: HIPPARCOS'97. Presentation of the 
  HIPPARCOS and TYCHO catalogues and first astrophysical results of the Hippaccos 
  space astrometry mission, B. Battrick, M. A. C. Perryman, \& P. L. Bernacca, 
  (eds.), ESA SP-402, Noordw\"{y}k, (1997), p.675 (JW).
\bibitem{} S. Bilir, Y. Karata\c{s}, \& S. G. Ak, \it TJPh, \bf 27, \rm (2003), 235. 
\bibitem{} H. J. Newberg, \& B. Yanny, \it ApJS, \bf 113, \rm (1997), 89.
\bibitem{} A. Th. Purgathofer, \it Bulletin/Lowell Observatory,  \bf 7, 
  \rm (1969), No: 147, p.98.
\bibitem{} R. Buser, in, Fresneau A., Hamm M. (eds.) Impacts des surveys du 
  visible sur notre connaissance de la Galaxie. Comptes rendus sur les Journees 
  de Strasbourg, 9eme Reunion, Obs. de Strasbourg, (1988), p.15. 
\bibitem{} D. J. Schlegel, D. P. Finkbeiner, \& M. Davis, \it ApJ, \bf 500, 
  \rm (1998) 525.
\bibitem{} D. Burnstein, \& C. Heiles, \it AJ, \bf 87, \rm (1982), 1165.
\bibitem{} R. P. Fenkart, \& S. Karaali, \it A\&AS,  \bf 69, \rm (1987), 33.
\bibitem{} W. Becker, \it ZA, \bf 54, \rm (1962), 155.
\bibitem{} W. Becker, \& Ch. Fang, \it A\&A, \bf 22, \rm (1973), 187.
\bibitem{} Th. Hersperger, \it A\&A, \bf 22, \rm (1973), 195.
\bibitem{} W. Becker, \it A\&AS, \bf 38, \rm (1979), 341. 
\bibitem{} G. Del Rio, \& R. P. Fenkart, \it A\&AS,  \bf 68, \rm (1987), 397.
\bibitem{} U. W. Steinlin, \it ZA, \bf 69, \rm (1968), 276.
\bibitem{} W. Gliese, \& H. Jahreiss, Third Catalogue of Nearby Stars
  (Preliminary Version), Astron. Rechen Inst. Heidelberg, (1992).
\bibitem{} R. Buser, \& U. Kaeser, \it A\&A, \bf 145, \rm(1985), 1.
\end{thebibliography}
\end{document}